# Laser-induced reversion of $\delta'$ precipitates in an Al-Li alloy: Study on temperature rise in pulsed laser atom probe

M. Khushaim [a*], R. Gemma [a*], T. Al-Kassab [+]

[a] King Abdullah University of Science and Technology (KAUST), Division of Physical Sciences and Engineering, Thuwal, Saudi Arabia 23955-6900.

**Abstract:**

The influence of tuning the laser energy during the analyses on the resulting microstructure in a specimen utilizing an ultra-fast laser assisted atom probe was demonstrated by a case study of a binary Al-Li alloy. The decomposition parameters, such as the size, number density, volume fraction and composition of $\delta'$ precipitates, were carefully monitored after each analysis. A simple model was employed to estimate the corresponding specimen temperature for each value of the laser energy. The results indicated that the corresponding temperatures for the laser energy in the range of 10 to 80 pJ are located inside the miscibility gap of the binary Al-Li phase diagram and fall into the metastable equilibrium field. In addition, the corresponding temperature for a laser energy of 100 pJ was in fairly good agreement with reported range of $\delta'$ solvus temperature, suggesting a result of reversion upon heating due to laser pulsing.

* Electronic address: Muna.Khushaim@kaust.edu.sa and Ryota.Gemma@kaust.edu.sa

[+] On leave, D-37077, Goettingen, Germany

# 1. Introduction:

Over the last 15 years, the APT technique has been improved significantly, making it a well-established nano-analysis tool in the field of material science (Blavette,1993). It has been extensively applied to the investigation of different types of materials due to its ability to map the distribution of single atoms in a material in real space on a nearly atomic scale; thus, using APT allows a better understanding of the physical phenomena involved in a material. The unique properties of this instrument have been used to study a wide variety of problems of current scientific and technological interest.

The time of flight (TOF) atom-probe technique was only applied to metallic samples due to the inability to transmit the nanosecond-scale, high-voltage pulses through a sample with a low electrical conductivity to its tip apex (Kellogg, 1980). Many attempts have been made to overcome these drawbacks of the atom probes. The primary concern was to extend the field of application of the technique to include poor conductivity materials. Efforts have also been made to increase the quantity of data resulting from a sample investigation by increasing the field of view (Gault, 2006). On this basis, the voltage pulses in a conventional atom probe were replaced by an ultra-short laser pulse, which is used to induce field evaporation (Kellogg, 1980). However, poor understanding of the laser interaction with the sample remains an issue.

To date, two models have generally been accepted on the role of the laser pulses. In the first model, a portion of the laser energy is absorbed by the specimen tip, leading to the generation of a thermal spike on the tip; thus, a temperature rise occurs. In this case, evaporation is described by the pure thermal-pulse model (TPM) proposed by Liu and Tsong (1984). In the second model, a strong electromagnetic field is expected at the sample surface due to the high instantaneous power of the laser and due to the enhancement effects related to the sub-wavelength dimensions of the sample (Martin, 1997). As a result, an ultrafast electric field at the tip apex is generated due to the rectification of the optical field at the specimen surface. This mechanism has been proposed by Vella et al. (2006). The pulsed field induced by optical rectification (OR) acts exactly as a voltage pulse with a sub-picosecond duration applied to the tip. Several results with pure metals, such as aluminum and tungsten, support the OR mechanism (Gault, 2006). Conversely, recent studies by Cerezo et al. (2007) have interpreted their experimental results with a pure TPM; they observed the surface diffusion of atoms caused by femtosecond laser irradiation (Cerezo, 2006). Vurpillot et al. (2009) measured the temperature of the specimen after a femtosecond laser pulse using a pump-probe method and demonstrated a temperature rise of approximately 600 K for tungsten. It has recently been shown that the interaction of the laser pulse with the tip leads to a complex absorption pattern (Koelling, 2011) and ultimately to a confined absorption of the laser energy in a hot spot with the dimensions of the laser wavelength (Houard, 2011); this suggests that the evaporation of ions during laser-assisted atom-probe tomography is a thermally driven process in which the tip apex is heated to a temperature that allows the thermal promotion of the field-evaporation process (Cerezo, 2006). Until recently, there was controversy over the physical mechanism of the field evaporation induced by a laser pulse, but Vurpillot et al. (2009) showed that the physical process behind field evaporation assisted by ultra-short laser pulses in metal is mainly described by the TPM model, declaring the influence of OR mechanism to be probably of a second order.

In this study, we investigate the details of the laser-induced specimen-heating effect using an interface reaction in an Al − Li alloy as a model system. This alloy is known to have a low-temperature, metastable, miscibility gap (Gayle, 1984). It has been shown that under classical conditions of ageing of this alloy, including solution treatment, fast quench to room temperature and thermal ageing at intermediate temperatures (e.g., 100-200 ˚C), the precipitation behavior was dominated by the presence of the metastable $\delta'(Al_3Li)$ phase with an $L1_2$ structure (Sanders, 1984). The influence of the laser power on the morphology, the composition and the diffusion of $\delta'(Al_3Li)$ precipitates in the aluminum-lithium-based alloy is identified. A simple model is used to explain the observed experimental behavior and to estimate the corresponding tip-apex temperature for various laser energies.

## 2. Materials and methods:

Ingots of Al – 2.0 wt.% (7.8 at.%) Li were drawn into 0.2-mm wires. The composition of the wires was measured via inductively coupled plasma optical emission spectrometry (Varian ICP-OES 720-ES). These wires were solution-treated at 500 ˚C for 30 min, followed by quenching in ice water. Then, the wires were aged at 190 ˚C for 3 h. This heat treatment is known to induce the formation of spherical $\delta'(Al_3Li)$ precipitates with $L1_2$ structure (Nozato, 1977). Atom probe tips were prepared via a standard electrochemical polishing procedure using a solution of 30 vol.% nitric acid in methanol at -20 ˚C. Atom probe tomography (APT) analyses were performed with both a CAMECA laser assisted wide angle tomographic atom probe (LAWATAP) for the field ion microscopy (FIM) mode and the CAMECA local electrode atom probe (LEAP 4000X HR). Data were acquired utilizing either the voltage- or the laser pulse mode. A diode-pumped (Nd:YAG) solid-state laser operating in the frequency-tripled ultraviolet region with a wavelength of 355 nm, a pulse duration of approximately 12 ps and a repetition rate of 200 kHz was used. The laser pulse energy was systematically varied through the following values: 10, 30, 40, 50, 60, 80 and 100 pJ. The position of the laser spot on the specimen was monitored using a charge-coupled device (CCD) camera. Data were acquired at a base temperature of 22 K and an average detection rate of 0.001 ions per pulse. The base pressure was maintained at less than $10^{-8}$ Pa during the analysis. The pulse fraction for the analysis in the voltage mode was set to 18%. Field ion images were obtained using Ne gas at $1.2 \times 10^{-3}$ Pa. Volume reconstruction of the acquired data was performed using the TAP3D and IVAS 3.6.6, software programs provided by CAMECA.

The reconstruction algorithm used was the standard evolution algorithm (Gault, 2012). The calibration of the primary reconstruction parameters was performed by observing the correct plane distance between the aluminum atomic planes for the respective crystallographic pole of (111) or (100) and the complete spherical shape of the precipitates. The value of the field strength of the evaporation $E_F$ was tuned to obtain the correct interplanar spacing of pure Al and the complete spherical shape of the precipitates. The effect of the solute Li on the $d$ spacing is considered to be negligible (Kellington, 1969). At 100 pJ, the atomic planes were not clearly resolved; thus, the reconstruction was performed based on the initial tip curvature radius measured via SEM and thereof corresponding $E_F$ (Vnm$^{-1}$).

To obtain the chemical composition within both the matrix and the precipitates, the linear composition profile along the cylinder perpendicular to the isoconcentration surface was used. Isoconcentration surfaces delineate the regions containing more than 14 at.% Li; this surface was obtained by sampling the atom probe tomography reconstruction with $1 \times 1 \times 1$ nm$^3$ voxels after applying a delocalization procedure that was developed by Hellman et al. (2003) with the smoothing parameters of 3 nm for the x- and y-coordinates and 1.5 nm for the z-coordinate.

Estimation of the size of the precipitates was performed using a cluster identification algorithm (Vaumousse, 2003) implemented in IVAS 3.6.6 software. Li-containing precipitates were identified using a maximum separation distance of 0.5 nm between Li atoms, and a minimum of 20 Li atoms in each precipitate.

Before and after the APT analyses, scanning electron microscopy (SEM) was used to observe the specimen apex geometry using Quanta 3D microscope (FEI).

## 3. Results and discussion:

In general, a reference measurement is used to investigate the potential influence of the laser pulse energy in an APT analysis; this can be done by performing an analysis in voltage-pulse mode. In this mode, a standing voltage $V_{dc}$ is applied to the specimen and then a short, additional voltage pulse will initiate the emission of the atoms. Figure 1 shows a three-dimensional reconstruction volume of the respective sample analyzed by the voltage pulse mode. At a glance, Li-enriched regions are clearly shown in Fig. 1 (a). This enrichment was found to have a nearly spherical shape. Concentration profiles were drawn across the cylinders perpendicular to these Li-enriched regions, and one typical shape of the profile is presented in Fig. 1(b) as an example. Within the enriched area, the Li concentration $c_{Li}$ reaches a maximum of 23 at.%, while in the matrix, the $c_{Li}$ is less than 6 at.%, suggesting a precipitation of a Li-rich phase. The average concentration of Li in the precipitate (calculations were performed on 15 precipitates) was determined to be $c_{Li} = (22.6 \pm 0.9)$ at.%; this value is similar to that of the $\delta'(Al_3Li)$ phase predicted for the Al-Li system at 190 ˚C (Hallstedt, 2007). The average diameter of the precipitates was estimated to be $(14.2 \pm 3)$ nm. This result is also in agreement with the results reported by Al-Kassab et al. (1991) and Krug et al. (2008), hinting at the precipitation of $\delta'(Al_3Li)$. All of the observations described above suggest that the Li-rich phase corresponds to metastable $\delta'(Al_3Li)$. TEM observations showed that this metastable phase is coherent with the Al matrix and has marginal coherency strains and small interfacial energy (Williams, 1975).

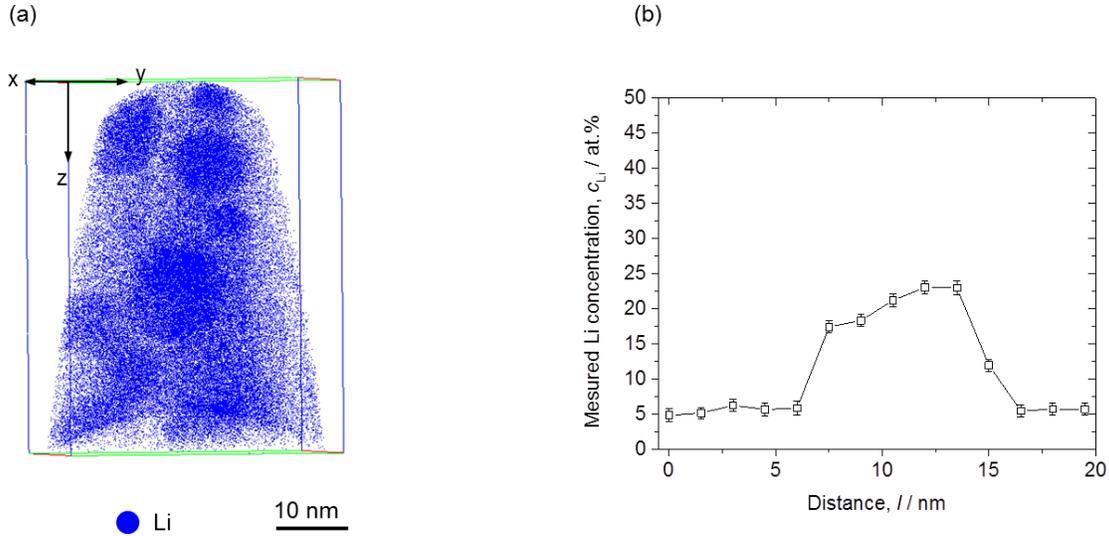

**Figure 1** Initial microstructure of the tip analyzed by the voltage mode. (a) The reconstructed volume showing the distribution of spherical $\delta'$ precipitates (Displayed atomic density = 20 at.% Li). (b) Linear composition profile along the cylinder with the size of (x, y = 11 nm and z = 20 nm) perpendicular to the isoconcentration surfaces delineate region containing 14 at. % Li. The average concentration of Li in the precipitate is $c_p = (22.6 \pm 0.9)$ at.% and in the matrix is $c_t = (5.4 \pm 0.8)$ at.%.

In laser-pulse mode during the APT analysis, a standing voltage $V_{dc}$ is applied to the specimen, and the pulsed field is generated by the laser itself. It is well known that the field strength of evaporation $E_F$ decreases with increasing laser energy, $E$ (Kellogg, 1980). This relationship, $E$ vs. $E_F$, has been confirmed by Figure 2, which shows the influence of the laser energy on the field strength of evaporation of the respective microstructure. This result confirms that field evaporation occurs at the effective field evaporation temperature $T_{eff}$ (Gault, 2010) during a pulsed-laser atom-probe procedure; this temperature is higher than the specimen base temperature. Field evaporation will be triggered at a lower electric field $F_{eff}$ than that required in the case of a conventional, high-voltage (HV) pulsed atom probe analysis of an identical specimen.

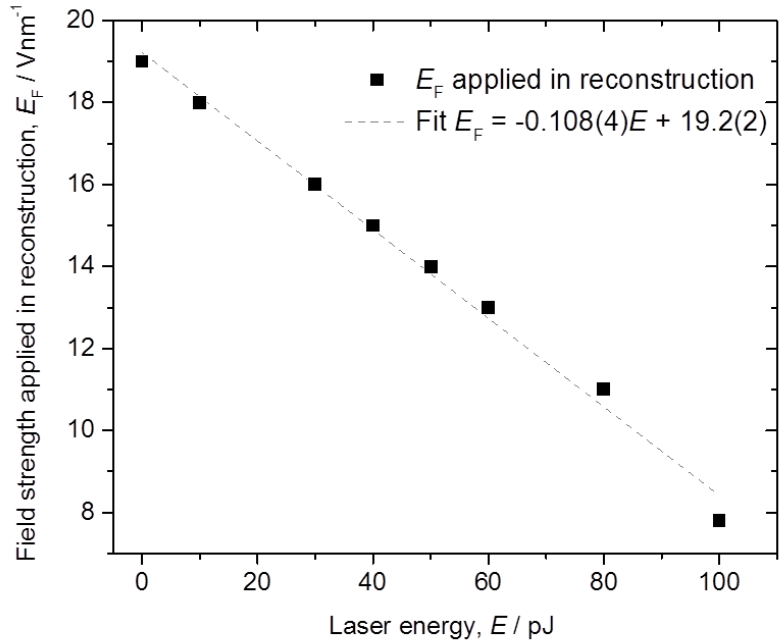

**Figure 2** Field strength of evaporation $E_F$ applied in reconstruction as a function of the laser energy $E$.

As a result of this thermal process, the morphology and composition of the $\delta'(Al_3Li)$ precipitates might be affected and presumably altered during the laser-mode analysis. Analyses at various laser energies were performed to monitor this effect, as shown in Figure 3, which shows a series of reconstructed volumes of the tips analyzed by laser pulses at the following laser energies: 10, 30, 40, 50, 60, 80 and 100 pJ. From Fig. 3(a-e), spherical precipitates in the microstructure are clearly visible in the range of laser energies from 10 to 60 pJ. Conversely, Fig.

3(f) shows that the precipitates begin to lose their distinctive shapes at 80 pJ. At an even higher energy of 100 pJ, precipitates are no longer detected as individual particles (Fig. 3(g)). Moreover, some enriched Li regions can be seen in the top view of the reconstructed volume in Fig. 3(h). Reproducibility of this absence of precipitates was confirmed 3 times over 4 experiments. The overall $c_{Li}$ in Fig. 3 (g, h) ranged from 5 to 19 at.% depending on the region over which the analysis was performed. It should be emphasized in this study that the reconstruction of the data was performed based on the adjustment of the interplanar spacing of (111) or (100) planes while assuming a hemispherical shape of the tip apex.

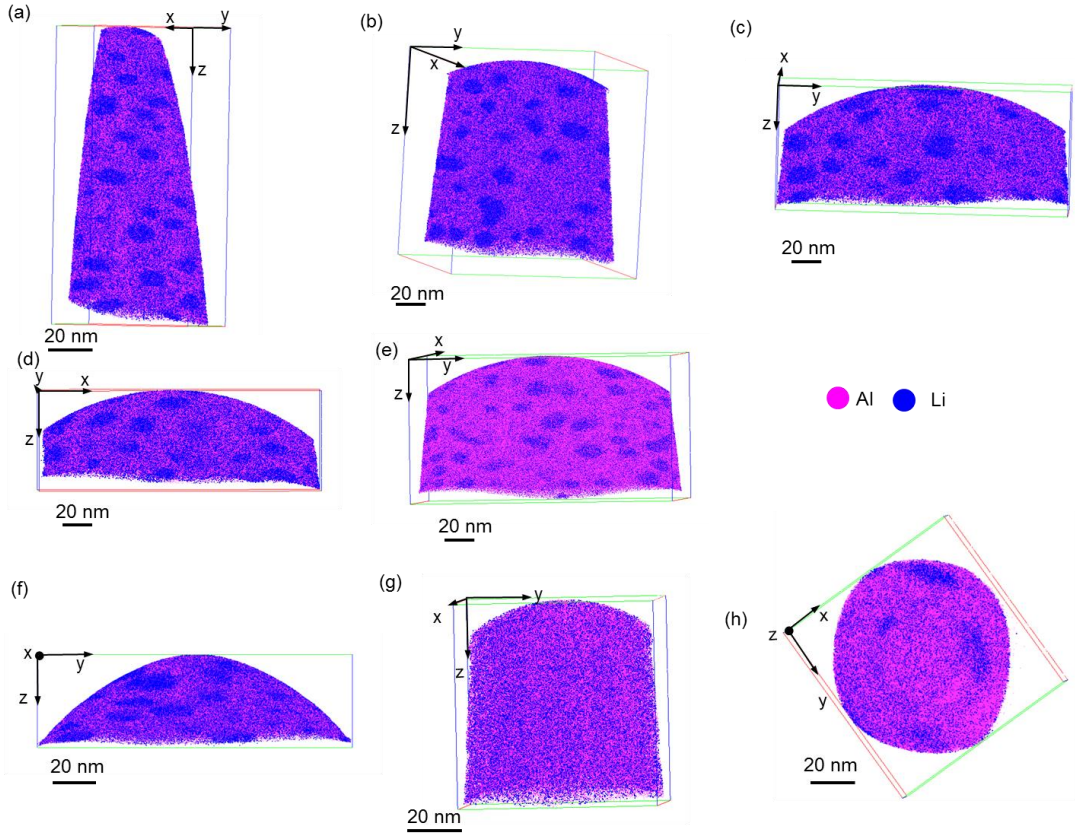

**Figure 3** Comparison of the morphology of the $\delta'$ precipitates under illumination of the laser pulses at various values of energy $E$ measured for different samples: (a) 10, (b) 30, (c) 40, (d) 50, (e) 60, (f) 80 and (g, h) 100 pJ (Displayed atomic density = 5.4 at.% Al, 20 at.% Li).

However, this assumption of a hemispherical tip shape is not always valid (Marquis, 2011). Laser irradiation on a tip is known to often cause tip reshaping, which results in erroneous 3D reconstructions (Shariq, 2009). Thus, SEM observations of the specimen apex geometry before and after the analysis at higher laser energies (i.e., 100 pJ) have been performed to describe the specimen's end form in this study. SEM micrographs of the tip end forms are compared in Figure 4(a, b). In both cases, the tip end form indeed resembles a nearly hemispherical shape. Consequently, at a laser energy of 100 pJ, the aforementioned tip-reshaping effect can be ignored. These observations justify the approach of using the conventional data reconstruction method in this study based on the assumption of a hemispherical end form of the specimen apex (Miller, 1996).

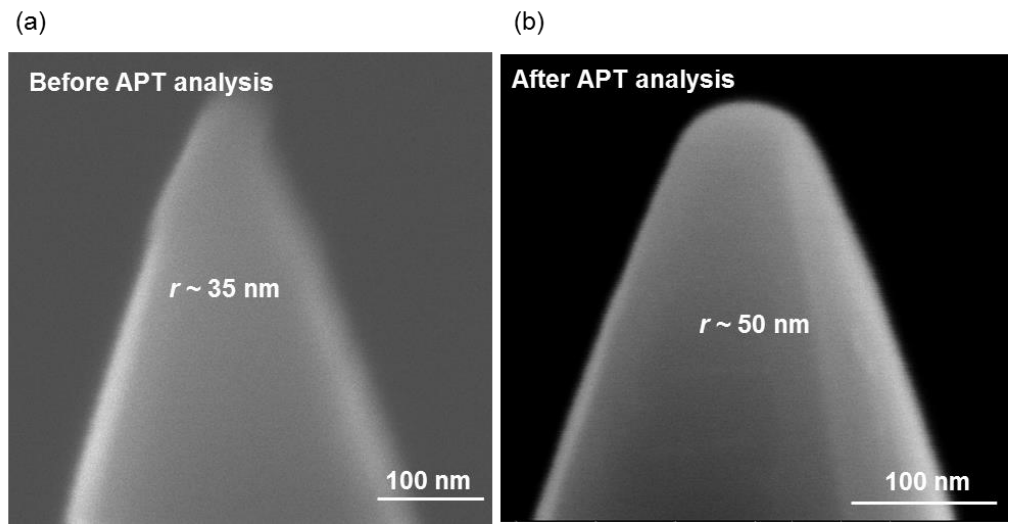

**Figure 4** Scanning electron micrographs showing the geometry of the tip apex (a) before and (b) after APT analysis by the laser pulse mode at 100 pJ.

FIM investigations were performed as well to verify the microstructure observed in the APT results and also to confirm the specimen's apex geometry after the analysis via the voltage-pulse and laser-pulse modes at the higher laser energies. It is known that the calculated field strengths of evaporation $E_F$ of Li and Al are 14 V/nm and 19 V/nm for single charged ions, respectively (Miller, 2000). Accordingly, it is likely that Li-rich regions undergo preferential evaporation of Li over Al in a FIM image; thus, the surface morphologies are modified as a result of a self-induced process during field evaporation. Consequently, after the field evaporation of the surface, preferentially evaporated regions are less likely to experience the ionization of the imaging gas compared to the rest of the surface and thus should be imaged with a dark contrast with respect to the matrix. This dark contrast was observed also by Hono et al. (1992). This contrast difference is also shown in this study, as identified by the circles in Figure 5(a). In combination with the APT results, it is thus reasonable to consider that these dark regions represent $\delta'(Al_3Li)$ precipitates, and the bright contrast region represents the matrix phase, which is enriched with Al. One bright spot appearing in the image is presumably due to the presence of oxide nuclei on the top most surface of the tip and is of minor importance to this study. Conversely, the image taken after the analysis at a laser energy of 100 pJ allows for completely different impression. No precipitates are visible; instead, a rather homogenous microstructure with different concentric rings that correspond to different crystallographic poles are observed (Fig. 5(b)). This is again consistent with the reconstructed volume in Fig. 3(g, h). As described earlier, the evolution of field evaporation in the FIM is associated with the self-modification of the surface geometry because the evaporation process is not controlled by other means to accommodate the different values of $E_F$ of the constituent elements. In the APT technique, evaporation is partially controlled by the superposition of the voltage and laser pulses over the DC voltage such that it minimizes the influence of the preferential evaporation; however, this influence is known to be significantly dependent on the pulse voltage (i.e., pulse fraction) and the pulse laser energy.

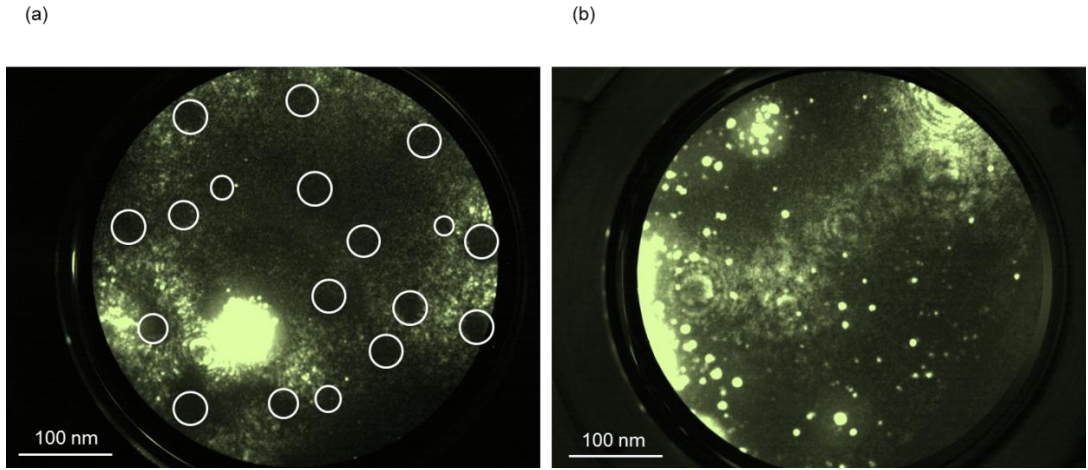

**Figure 5** Field ion micrographs of the aluminum lithium specimen (a) Recorded for the specimen after analysis by the voltage pulse mode, shows the distribution of darkly imaged $\delta'(Al_3Li)$ precipitates as marked by circles. (b) Recorded for the specimen after analysis by the laser pulse mode at the laser energy $E = 100$ pJ, shows a homogenous microstructure, with different concentric rings corresponding to different crystallographic poles.

Thus, the measured Li concentrations $c_{Li}$ in the proposed samples were carefully examined in both the voltage-pulse and laser-pulse modes at different laser energies. In Figure 6, $c_{Li}$ is plotted as a function of the laser energy. In voltage mode ($E = 0$ pJ), the detected $c_{Li}$ was $(5.4 \pm 0.8)$ at.% in the matrix, which is somewhat lower than the expected solubility limit of Li at 190˚C (6 at.%), as reported by Noble and Bray (1998); this implies a marginal influence from the preferential evaporation of Li. However, the error bar verifies that the detected $c_{Li}$ is within the expected range. Similarly, $c_{Li}$ in the precipitates falls within a reasonable range of $(22.6 \pm 0.9)$ at.%, closely matching the reported value (Hallstedt, 2007). It is therefore concluded that the proposed APT measurement is not significantly affected by the mentioned preferential evaporation of Li within the detectable level. However, as the laser energy is increased, a significant departure from those values is observed; $c_{Li}$ in the precipitates' $c_p$ drops to 19.5 at.%, while $c_{Li}$ in the matrix' $c_t$ increases to 7.6 at.%. When considering the expected temperature rise induced by laser irradiation, this trade off relation between $c_p$ and $c_t$ reminds us of the lever rule at the miscibility gap.

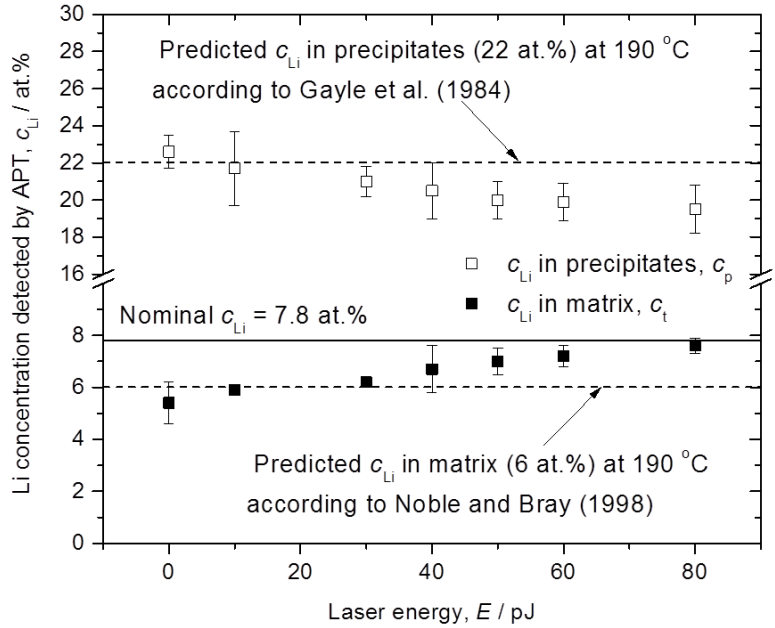

**Figure 6** Measured matrix Li concentrations $c_{Li}$ in the tip analyzed by the voltage pulse mode and by the laser pulse mode at different values of laser energy as a function of the laser energies $E$ (in voltage mode, $E = 0$ pJ).

To describe the origin of this change in $c_{Li}$, the precipitates shown in Figs. 1 and 3 were further characterized to obtain quantitative information on the precipitates' diameters, their number densities and their volume fractions. Using the simple law of mass conservation, a relationship for the spherical precipitates between the radius $r$, the number density $N_v$ and the volume fraction $f$ is given by the following relationship (Balluffi, 2005):

$$^{4\pi}/_3 N_v r^3 = f \qquad (1)$$

Additionally, the lever rule holds to the following relationship (Balluffi, 2005):

$$^{c_0 - c_t}/_{c_p - c_t} = f \qquad (2)$$

Where $c_0$ is the nominal Li concentration in the alloy, $c_p$ is the measured Li concentration within the precipitates and $c_t$ is the measured Li concentration within the matrix.

The methodologies described in materials and methods section are used to extract information about the average compositions and the average diameters of the precipitates in all of the reconstructed volumes in Figs. 1 and 3. For using Eq. (1), the total analyzed volume has been calculated from the atomic volume for each ion multiplied by the total number of the ions in the analyzed volume. By dividing the number of precipitates by the calculated value of the volume of analysis, an approximate number density of the precipitates was obtained.

A summary is shown in Table 1 as a list of the average diameter $d$ of the precipitates, their number density $N_v$, their volume fraction $f$, their average Li compositions within the precipitates $c_p$ and the Li composition of the matrix $c_t$. Data for 100 pJ is not available due to the lack of precipitates in the volume analyzed, as shown in Fig. 3 (g, h). Noticeably, the values of $d$, $N_v$ and $f$ decrease as the laser energy increases from 10 to 80 pJ. Eq. (2) was used to confirm the validity of $f$ estimation from Eq. (1). The difference between these two estimations by Eqs. (1) and (2) of $f$ was ±1% at the largest and is thus negligible. Strictly speaking, Eq. (2) holds under equilibrium and might not directly account for the metastable miscibility gap, when other stable precipitates such as $\delta(AlLi)$ phase are formed. However, it has been shown that the $\delta'(Al_3Li)$ phase is formed and accounts for most of the precipitates in the Al − 2.0 wt.% Li alloy aged at 200 ˚C within ageing times up to 240 h (Nozato, 1977). This is in a good agreement with the proposed experimental observation; we did not detect phases other than $\delta'(Al_3Li)$; thus, mass conservation is believed to hold between the $\alpha$ (Al matrix) and $\delta'$ phases. Therefore, if the change in Li concentrations (i.e., $c_p$ and $c_t$ in Fig. 6) is associated with the temperature increase due to laser irradiation, the proposed data should follow the suggested metastable miscibility gap of $\alpha + \delta'$ in Al-Li system (Gayle, 1984). Conversely, $d$, $N_v$ and $f$ were found to decrease with increasing laser energy and thus increasing temperature, which is opposite to the conventional kinetic model of coarsening, which was proposed as the Lifshitz-Slyozov-Wanger (LSW) theory (Jayanth, 1989).

**Table 1** Summary of the data obtained for the diameter $d$, number density $N_v$, volume fraction $f$, Li composition $c_p$ in the precipitates and Li composition $c_t$ in the matrix for the sample analyzed both in the voltage mode and laser mode at various values of laser energy. $f$ was calculated by Eq. (1).

|         | Diameter $d \times 10^{-9}$ (m) | Number density $N_v \times 10^{22}$ (m$^{-3}$) | Volume fraction $f$ % | Li concentration in precipitates $c_p$ (at.%) | Li concentration in matrix $c_t$ (at.%) |
|---------|---|---|---|---|---|
| **Voltage** | 14.2 ± 3 | 10 ± 0.1 | 14 ± 2 | 22.6 ± 0.9 | 5.4 ± 0.8 |
| **L 10 pJ** | 13.4 ± 3.1 | 9.68 ± 0.5 | 12 ± 1 | 21.7 ± 2 | 5.9 ± 0.1 |
| **L 30 pJ** | 12.8 ± 3.3 | 9.23 ± 0.1 | 10 ± 1 | 21 ± 0.8 | 6.2 ± 0.2 |
| **L 40 pJ** | 12.4 ± 3.1 | 9.18 ± 0.01 | 9 ± 0.1 | 20.5 ± 1.5 | 6.5 ± 0.9 |
| **L 50 pJ** | 11.5 ± 3.4 | 8.39 ± 0.5 | 6 ± 0.1 | 20 ± 1 | 7 ± 0.4 |
| **L 60 pJ** | 10.6 ± 2.6 | 8.27 ± 0.01 | 5 ± 0.1 | 19.9 ± 1 | 7.2 ± 0.4 |
| **L 80 pJ** | 9.2 ± 1.5 | 6.25 ± 0.2 | 2 ± 0.5 | 19.5 ± 1.3 | 7.6 ± 0.3 |

Our experimental observations above anticipate that the $\delta'(Al_3Li)$ precipitates at various laser energies might involve an evolution of the tip temperature upon irradiation with the laser pulses. However, the extent of the real temperature increase is not yet known, and to date, we cannot compare the results of this study with the Al-Li phase diagram. We therefore attempt to estimate the tip temperature $T$ as described in the following paragraph.

The diffusion lengths $L$ and effective diffusion coefficients $D$ were calculated for the Li atoms from the $\delta'$ precipitates at each laser energy by considering changes in the precipitates' diameter $\Delta d$ as being equal to the diffusion length $L$. The values $\Delta d$ have been estimated from the difference between the diameter of the precipitate at each laser energy and their original diameter before applying the laser pulses (i.e., in voltage mode). If the reaction is assumed to be diffusion-controlled, the effective diffusion coefficients $D$ are estimated from the relationship between the diffusion length $L$ and the diffusion time $t$ (Balluffi, 2005):

$$\Delta d \cong L = 2\sqrt{Dt} \qquad (3)$$

$$D = L^2/4t \qquad (4)$$

As an approximate of the diffusion time $t$, we assume a diffusion time of 1 second in this study due to the high activation energy of the Li atoms in $\delta'(Al_3Li)$, which is $Q = 119$ kJmol$^{-1}$(Murch, 1990), and due to the study of kinetics of reversion by Okuda et al.(1993). In addition, this rough estimate of the diffusion time is based on an estimate of the diffusion penetration distance from a point source. Suppose a $\delta'$ particle is a point source of a diffusing substance and $L$ in Eqs. 3 and 4 is the diffusion penetration distance. Then, using eq. 4, the time required for the Li atoms to diffuse from the $\delta'$ particle through a distance of 1 nm in the microstructure is approximately 2.5 s. For simplicity, the diffusion time is set to 1 s due to the variety of precipitate sizes and, therefore, of diffusion lengths for different laser irradiation energies. Assuming that the previously reported value of $Q$ is also valid in the sample of the proposed study, and the expected reaction at the precipitate/matrix interface is one elemental process, as proposed by Okuda et al. (1993), the diffusion behavior of Li considered in this study is described by an Arrhenius-type relationship (Balluffi, 2005) :

$$D = D_0 \exp(-Q/RT) \qquad (5)$$

Where $D$ is the diffusion coefficient ($m^2 s^{-1}$), $D_0$ is the pre-exponential factor ($m^2 s^{-1}$), $Q$ is the activation energy ($kJ mol^{-1}$), $R$ is the gas constant ($J mol\ K^{-1}$) and $T$ is the thermodynamic temperature (K).

Using Eqs. (4) and (5), $T$ at each laser energy was estimated. The diffusion lengths $L$ and the effective diffusion coefficients $D$ for the Li atoms at each laser energy are summarized with their corresponding temperatures $T$ in Table 2. Based on the table, the diffusion length $L$ increases with increasing laser energy and thus temperature. As a result, the estimated diffusion coefficient $D$ scales with the temperature $T$. This relationship is more evident in Figure 7. These data fall into a reasonable range of $D$ compared to that of the bulk Al-Li system, where $E = 10 - 80$ pJ. At 100 pJ, only the lower limit of $D$ and $T$ can be estimated by considering the complete re-dissolution of the precipitates. Based on this crude estimation, the $T$ at 100 pJ is at least 540 K (267 °C). The real $T$ at 100 pJ can be substantially higher than this value. This margin is indicated by the shaded region in Fig. 7. The estimated $D$ in this study could be, however, still somewhat underestimated because the diffusion process of Li might be influenced by the continuous formation of a new surface during the APT analysis itself. This influence is often observed as a surface segregation in the APT analysis of a light element such as hydrogen or deuterium (Gemma, 2009). In this study, such segregation was not experimentally observed. The influence of tensile stresses caused by the DC electric field can also not be ruled out, but its contribution should not be significant based on the voltage APT data.

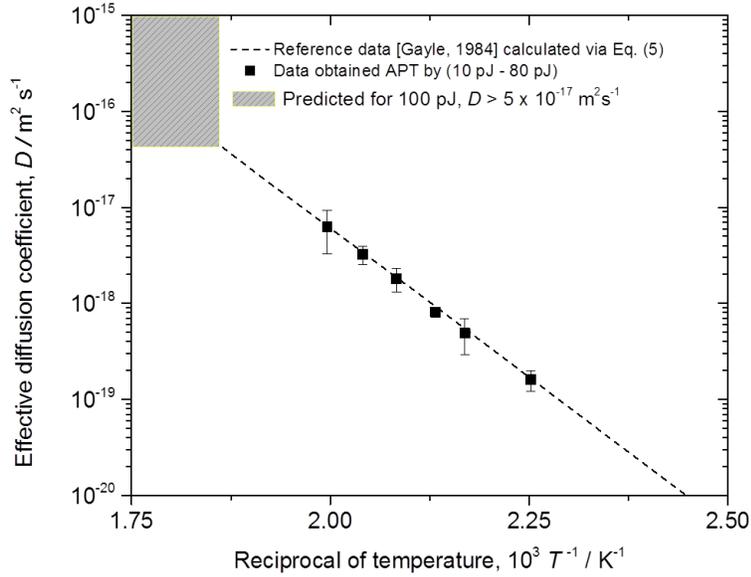

**Figure 7** Arrhenius plot of effective diffusion coefficients $D$ estimated by using Eqs. (4) and (5). Estimated values are compared with $D$ values of the bulk Al-Li system (Gayle, 1984) (broken line) calculated via Eq. (5).

In Fig. 2, it was shown that the field strength of evaporation $E_F$ monotonically decreases with increasing laser energy $E$. Additionally, the temperature rise $dT$ due to laser irradiation was suggested to be nearly proportional to the laser energy $E$ (Vurpillot, 2009). Thus, it is anticipated that the relationship of $E_F$ and $dT$ is inversely proportional; this seems to also be valid in the case investigated in this study within the range of 10 to 80 pJ, as shown in Figure 8. In this figure, if the linear relationship of $dT$ vs. $E_F$ is extrapolated to the $E_F$ of 100 pJ (7.8 Vnm$^{-1}$), the corresponding $dT$ produces a temperature of 511 ± 4 K; thus, this $dT$ plus the base temperature 22 K suggests a specimen temperature $T$ of 533 ± 4 K (260 °C ± 4). This value is not significantly different from 540 K (267 °C), which was separately deduced from the estimated diffusion coefficient via the precipitates' re-dissolution at 100 pJ (see Table 2 and Fig. 7). As described earlier, the estimation of $T$ at 100 pJ via the estimation of $D$ might not be simple; however, both approaches from $D$ and $E_F$ to estimate $T$ at 100 pJ produce comparable values of $T$ (533-540 K, 260-267 °C), which implies that a complete re-dissolution of $\delta'$ occurs at this temperature range.

**Table 2** The calculated diffusion lengths and effective diffusion coefficients for Li atoms from the precipitates and the estimated temperatures corresponding to each point of the laser energy.

|  | Diffusion length $L \times 10^{-9}$ (m) | Diffusion coefficient $D \times 10^{-19}$ (m$^2$s$^{-1}$) | Temperature $T$ (K) | Temperature $T$ (°C) |
|---|---|---|---|---|
| **L 10 pJ** | 0.8 ± 0.1 | 1.6 ± 0.4 | 444.4 ± 3 | 171.4 ± 3 |
| **L 30 pJ** | 1.4 ± 0.3 | 4.9 ± 2 | 460.5 ± 6 | 187.5 ± 6 |
| **L 40 pJ** | 1.8 ± 0.1 | 8.1 ± 1 | 468 ± 2 | 195 ± 2 |
| **L 50 pJ** | 2.7 ± 0.4 | 18 ± 5 | 480.6 ± 5 | 207.6 ± 5 |
| **L 60 pJ** | 3.6 ± 0.4 | 32.4 ± 7 | 490.3 ± 4 | 217.3 ± 4 |
| **L 80 pJ** | 5 ± 1.5 | 62.5 ± 30 | 501.6 ± 10 | 228.6 ± 10 |
| **L 100 pJ** | > 14.2 | > 521 | > 540 | > 267 |

Now, we attempt to compare the proposed values of $c_p$, $c_t$ and $T$ obtained via the APT technique with the metastable miscibility gap of $\alpha + \delta'$ in the Al-Li system that was previously summarized by Gayle et al. (1984) in Figure 9. The measured data points are located nearly within the miscibility gap of the diagram. The Li solubility in this study is shown to be in better agreement with the report of Noble and Bray (1998) compared to that of Gayle et al. (1984) (see data for 10 pJ in Fig. 9). It is also clear that the average Li concentrations in the $\alpha$ and $\delta'$ precipitates follow the gap closure of the miscibility gap with the temperature as the laser energy is increased up to 80 pJ; this behavior is supported by considering the rise in $T$ due to the laser irradiation at the sample tip.

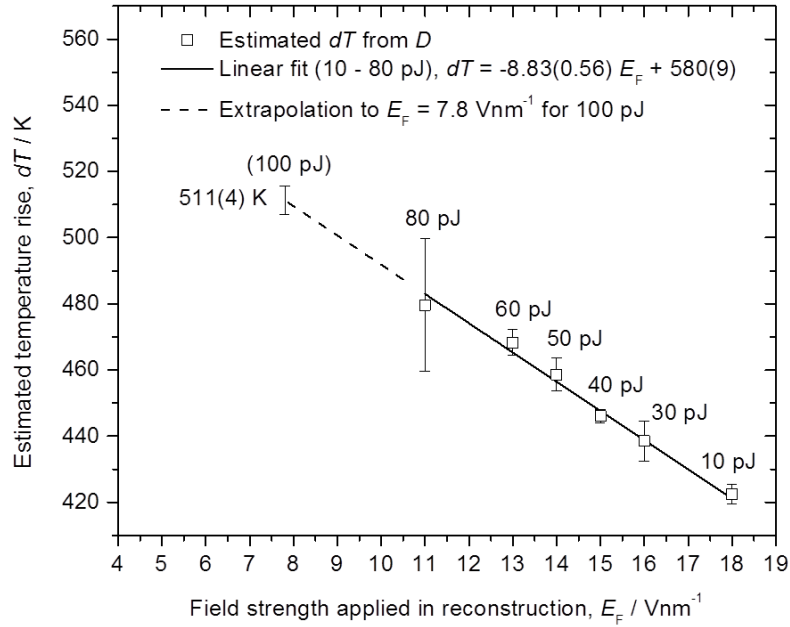

**Figure 8** Laser-induced specimen temperature rise $dT$ plotted versus electrical field strength of evaporation $E_F$. Inversely proportional relationship between $E_F$ and $dT$ is shown (At 100 pJ, $dT$ is extrapolated from the linear relationship due to the lack of information on diffusion length $L$.).

Judging from the results in Fig. 3 and Table 1, the diameter $d$ of the precipitates decreases with increasing laser energy. This does not correspond to coarsening but rather suggests a partial reversion of the precipitates at 10-80 pJ. The data at 10-80 pJ coincide with the miscibility gap, which is well below the critical temperature $T_c$; thus, we emphasize here using Fig. 6 that the Li concentration in the matrix measured in voltage mode and at 10 pJ nearly verifies the solubility of 6 at.% at 190°C, at which the aging treatment was performed. Departures from this solubility begin at laser energies above 30 pJ, where the corresponding specimen temperature exceeds 190˚C. This experimental behavior indicates that the increase in temperature as the laser energy increases is sufficiently rapid to destabilize all of the precipitates present in the initial microstructure, leading to a supersaturated $c_{Li}$ in the matrix, as shown in Fig. 6. As noted by Noble and Bray (1998), this supersaturation indicates the progressive dissolution of $\delta'$ precipitates. The temperature of the $\alpha/\delta'$ solvus in the Al – 2.0 wt.% Li alloy has been determined to be near $T_c$ = 513 K (240 ˚C) (Noble, 1998). Based on this value, the results at a laser energy of 100 pJ corresponds to full reversion, where no clear evidence of precipitation was detected in the FIM observation (Fig. 5(b)) or the APT analysis (Fig. 3 (g, h)). The corresponding temperature at 100 pJ was suggested to be in the range of $T$ = 533-540 K (260-267 ˚C) (Figs. 7, 8). This range of $T$ is in agreement with the reported range of the solvus temperature of $\delta'$ in the Al – 2.0 wt.% Li alloy: 513-523 K (240-250 ˚C) (Nozato, 1977) (Noble, 1998). Within this temperature range, the precipitates are nearly fully reverted into the matrix. This influence of the heat on $\delta'$(Al$_3$Li) precipitates was further confirmed by conducting a voltage-mode analysis immediately after the laser-pulse mode analysis at 100 pJ. Figure 10 shows the reconstructed volumes of an identical tip analyzed using, first, the voltage pulse mode; second, the laser pulse mode at 100 pJ; and finally, the voltage pulse mode. The top view of the reconstructed volume of the tip analyzed using the voltage pulse mode shown in Fig. 10 (a) confirms the initial stage of the microstructure. Spherical $\delta'$ precipitates are present around the (111) crystallographic pole. Irradiating the tip using a laser with a pulse energy of 100 pJ results in solute-enriched regions without spherical $\delta'$ precipitates in the Al matrix; these are shown in the top view of the reconstructed volume provided in Fig. 10 (b). This observation is similar to the result shown in Fig. 3 (h). The results of a voltage pulse mode analysis performed immediately after the laser pulse mode analysis whose results are shown in Fig. 10 (c) are the same as those shown in Fig. 10 (b). No spherical precipitates are detected; instead, some Li-enriched regions are visible. The depth concentrations measured along the cylinders perpendicular to the Li-enriched regions are shown in Fig. 10 (d, e). Fig. 10 (d) shows the depth concentration profile of the Li-enriched region shown in Fig. 10 (b). Once again, the overall $c_{Li}$ ranges from 5 to 19 at.%. The depth concentration profile of the Li-enriched region shown in Fig. 10 (c) is shown in Fig. 10 (e). The overall $c_{Li}$ ranges from 6 to 16 at.%. Variations in the shape and composition of the $\delta'$ precipitates after irradiation by a laser with a pulse energy of 100 pJ confirm the effect on the local microstructure of the specimen at the apex of the tip. These results suggest that irradiation by a laser pulse energy of 100 pJ has a permanent effect on the microstructure of the specimen at least locally at the tip apex. The disordering of $\delta'$ precipitates is also shown to not cause the dispersion of solute Li atoms into the matrix but rather leaves solute-enriched and disordered regions within the matrix (Nozato, 1977). The observations above support the proposed interpretation of the reversion phenomenon observed at 100 pJ. The results in Fig. 8 indicate that the range of $dT$ is between 422 and 518 K; these values are not significantly different from the reported $dT$ of 600 ± 50 K, as observed with tungsten using an infrared (IR) laser ($\lambda$ = 1030 nm) (Vurpillot, 2009) despite the fact that the laser wavelength used in that study was different. Based on the study by Houard et al. (2011), UV laser illumination on an Al or steel tip produces a localized energy absorption near the tip apex to a depth of 500 nm in the material, while the IR laser does not. Therefore, it would be reasonable to consider that the proposed case is the former (i.e., the heating zone could be localized near the tip apex). Thus, if the voltage-mode analysis conducted immediately after the laser mode runs through to a sample depth (e.g., 500 nm to 1 μm), the original microstructure of the $\delta'(Al_3Li)$, which is free from the influence of heat, might be progressively detected.

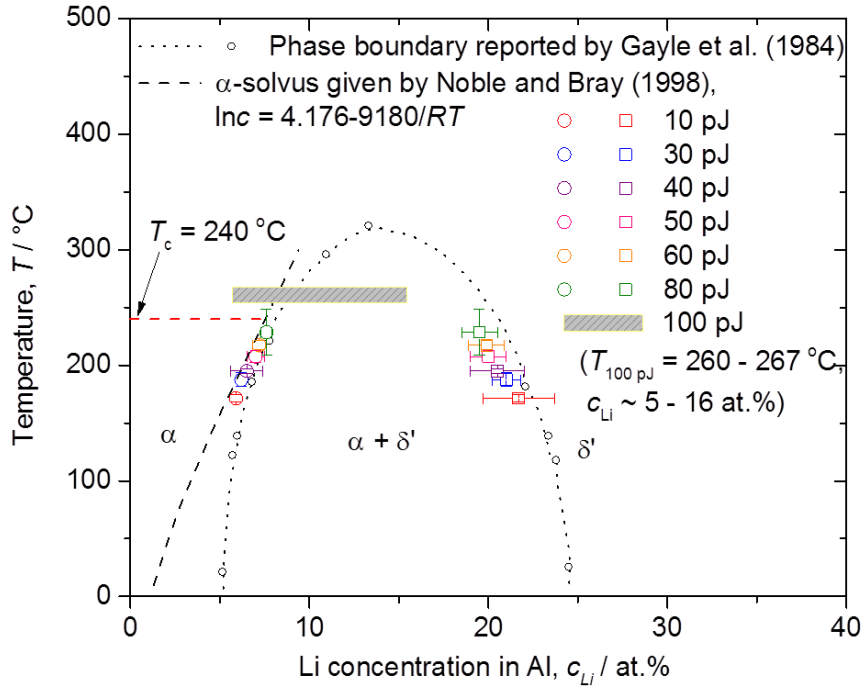

**Figure 9** Comparison of the experimental Li concentrations and the estimated temperatures with an Al-Li phase diagram adapted from (Gayle, 1984), (Noble, 1998). The data obtained at $E = 10 - 80$ pJ agree with the miscibility gap, well below the $T_c$. At 100 pJ, the tip temperature is estimated to be in the range of solvus temperature of $\delta'$ precipitates.

Regarding the thermal properties of the Al-Li system, it is known that the thermal conductivity decreases from $(2.50 \pm 0.04)$ Wcm$^{-1}$K$^{-1}$ to $(0.91 \pm 0.02)$ Wcm$^{-1}$K$^{-1}$ with increasing Li concentration in Al from 0 to 2 wt.% in the temperature range of 127 to 185 °C (Mcdonell, 1989). When considering higher Li contents within the precipitates than that of the matrix, the thermal conductivity of the respective microstructure would be even lower, leading to additional heating. Interfaces between these nano-sized precipitates and the matrix would effectively act to scatter phonons (Kim, 2006), which also suggests an increasing temperature. Hence, due to these microstructural phenomena, a direct comparison of the *dT* observed in a microstructure containing hetero- interfaces with that of pure metals is only conditional valid.

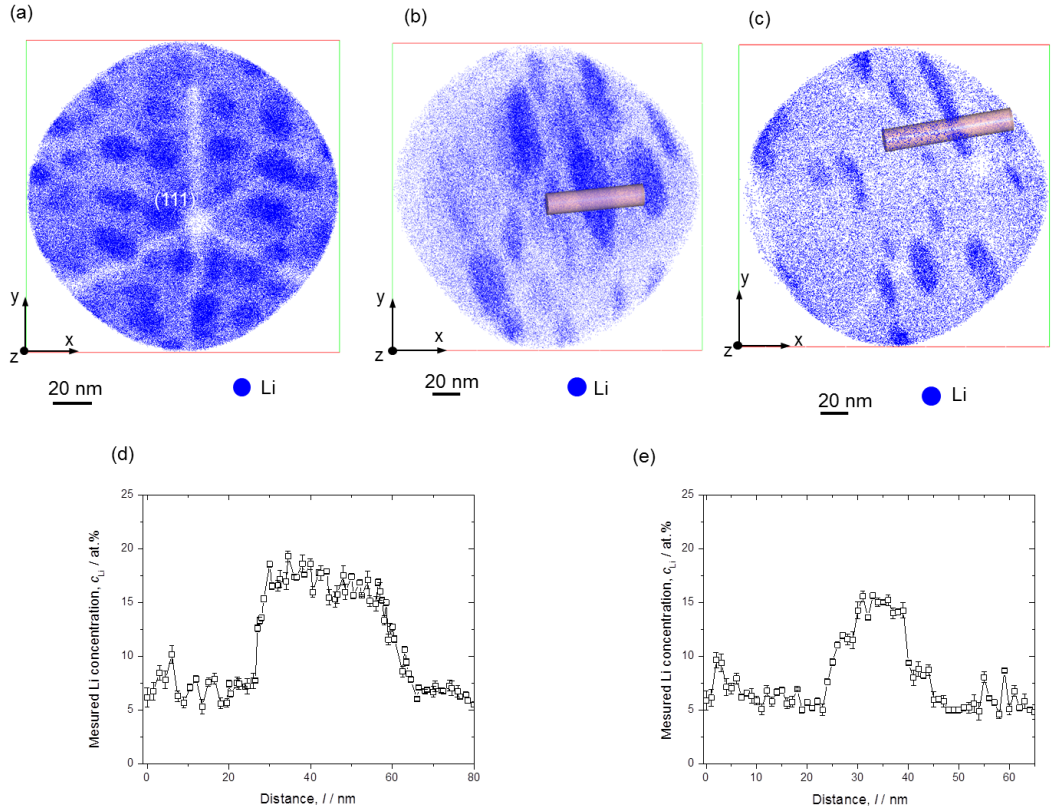

**Figure 10** Comparison of the morphology and composition of $\delta'$ precipitates before and after laser analysis: (a) Top view of the reconstructed volume of the tip analyzed by voltage pulse mode (Displayed atomic density = 30 at.% Li). (b) Top view of the reconstructed volume of the tip analyzed by laser pulse mode at 100 pJ (Displayed atomic density = 80 at.% Li). (c) Top view of the reconstructed volume of the tip analyzed by voltage mode after conducting laser analysis showing the permeant effect of the heating on the $\delta'$ precipitates (Displayed atomic density = 100 at.% Li). (d) Composition profile along the cylinder with the size of (x, y = 15 nm and z = 80 nm) perpendicular to Li enriched regions in (b). (e) Composition profile along the cylinder with the size of (x, y = 10 nm and z = 65 nm) perpendicular to Li enriched regions in (c).

In this study, atom probe tomography presented a series of snapshots during *in-situ* reversion of $\delta'(Al_3Li)$ precipitates, initiated by laser irradiation, using different laser energies for the first time. In addition, the attempt shown in this study might provide a method to investigate real sample temperatures during laser-APT analyses using an interface reaction itself as a probe.

## 4. Conclusion:

In this study, we report on the influence of the laser-pulse energy on the morphology and composition of the $\delta'(Al_3Li)$ precipitates in a binary Al-Li alloy. The influence of ultrafast laser pulses on the microstructures of the materials was investigated in terms of the variations in the diameter, $d$; the number density, $N_v$; the volume fraction, $f$; and the composition $c_\text{p}$ of the precipitates. The temperature increase of the sample via UV laser illumination was investigated by estimating the diffusion length and the diffusion coefficient of Li atoms. A comparison with the Al-Li phase diagram was performed. At a lower laser pulse power, all data were found to be within the miscibility gap of the diagram. However, depending on the laser energy, variations in the Li concentration were clearly detected along the solubility line of the precipitates. At the highest laser energy applied in this study (i.e., 100 pJ), the estimated tip temperature was near the critical temperature of $\delta'(Al_3Li)$ in the Al-2.0 wt.% Li alloy; this is supported by the results, which showed no precipitates in the analysis at this laser energy, which can be attributed to a significant and permanent heating effect causing a reversion of the once-formed $\delta'(Al_3Li)$ at the local region of the tip apex. In the course of the systematic investigation of the microstructural changes in the $\delta'(Al_3Li)$ that were caused by different laser pulse energies, this study proposed one simple method to estimate real specimen temperatures via laser-induced interface reactions during a laser-APT analysis.

**Acknowledgements:**

Prof. Alexander Rothenberger, Associate Professor of Chemical Science – King Abdullah University of Science and Technology is gratefully acknowledged. M. Khushaim gratefully acknowledges financial support provided through King Abdullah University of Science and Technology (KAUST) base-line funding program.

**References**:

Al-Kassab T, Menand A, Chambreland S, and Hassen P. 1991. The early stages of decomposition of Al-Li alloy. Surface Science 266: 333-336.

Blavette D, Bostel B, Deconihant B and Menand A. 1993. An atom probe for three-dimensional tomography. Nature 363(6428): 432-435.

Balluffi RW, Allen SM and Carter WC. 2005. Kinetics of Materials. WILEY INTERSCIENCE.

Cerezo A, Smith GDW and Clifton pH. 2006. Measurement of temperature rises in the femtosecond laser pulsed three-dimensional atom probe. Applied physics letters 88(15).

Cerezo A, Clifton PH, Comberg A and Smith GDW. 2007. Aspects of the performance of a femtosecond laser-pulsed 3-dimensional atom probe. Ultramicroscopy 107(9): 720-725.

Gault B, Vurpillot F, Vella A, Gilbert M, Menand A, Blavette D and Deconinhaut B. 2006. Design of a femtosecond laser assisted tomographic atom probe. Review of Scince Instrument 77(043705).

Gault B, Muller M, Fontaine ALa, Moody MP, Sharig A, Cerezo A, Ringer SP and Smoth GDW. 2010. Influence of surface migration on the spatial resolution of pulsed laser atom probe tomography. Journal of Applied physics 108(044904).

Gault B, Moody MP, Cairney JM and Ringer R. 2012. Atom Probe Microscopy. New York: Springer Series in Materials Science.

Gayle F, Vander Sande J and McAlister AJ. 1984. The Al-Li (Aluminum-Lithium) system. Bulletin of Alloy Phase Diagrams 5(1): 19-20.

Gemma R, Al-Kassab T, Kirchheim R, and Pundt A. 2009. APT analyses of deuterium-loaded Fe/V multi-layered films. Ultramicroscopy 109: 631-636.

Hallstedt B and Kim O. 2007. Thermodynamic Assessment of the Al-Li System. International journal of Metarial Research 98(10): 961-969.

Hellman OC, Du Rivage JB and Seidman DN. 2003. Efficient sampling for three-dimensional atom probe microscopy data, Ultramicroscopy 95: 199-205.

Hono K, Babu SS, Hiraga K, Okano R and Sakurai T. 1992. Atom probe study of early stage phase decomposition in an Al-7.8 at.% Li alloy. Acta Metallurgica et Materialia 40(11): 3027-3034.

Houar J, Vella A, Vurpillot F and Deconihout B. 2011. Three-dimensional thermal response of a metal subwavelength tip under femtosecond laser illumination. Physical Review 84(3): 033405.

Jayanth CS and Nash P. 1989. Factors affecting particle-coarsening kinetics and size distribution. Journal of Materials Science 24(9): 3041-3052.

Kellington SH, Loveridge D and Titman JM. 1969. The lattice parameters of some alloys of lithium. Journal of Applied Physics: D 2.

Kellogg GL and Tsong TT. 1980. Pulsed-laser atom-probe field-ion microscopy. Journal of Applied physics 51(2): 1184-1193.

Koelling S, Innocenti N, Hellings G, Gillbert M, Kambham AK, De Meyer K and Vandervorst W. 2011. Characteristics of cross-sectional atom probe analysis on semiconductor structures. Ultramicroscopy 111(6): 540-545.

Kim W, Zide J, Gossard A, Klenov D, Stemmer S, Shakouri A and Majumdar A. 2006. Thermal Conductivity Reduction and Thermoelectric Figure of Merit Increase by Embedding Nanoparticles in Crystalline Semiconductors. Physical Review Letters 96(4): 045901.

Krug ME, Dunand DC and Seidman DN. 2008. Composition profiles within Al3Li and Al3Sc⁄Al3Li nanoscale precipitates in aluminum. Applied physics letters 92(12): 124107.

Liu F, and Tsong TT. 1984. Numerical calculation of the temperature evolution and profile of the field ion emitter in the pulsed-laser time-of-flight atom probe. Review of Science Instrument 55(1779).

Marquis EA, Geiser BP, Prosa TJ and Larson DJ. 2011. Evolution of tip shape during field evaporation of complex multilayer structures. Journal of Microscopy 241(3): 225-233.

Martin OJ and Girard C. 1997. Controlling and tuning strong optical field gradients at a local probe microscope tip apex. Applied physics letters 70(6): 705-707.

Mcdonell WR, Albensius EL and Benjamin RW. 1989. Aluminum-Lithuim target behaviour. in Report WSRC – RC – 89 – 970. Westinghouse Savannah River Company: Savannah River Site Aiken SC USA.

MillerMK, Cerezo A, Hetherington MG and Smith GDW. 1996. Atom Probe Field-Ion Microscopy. New York: Oxford University Press.

Miller MK. 2000. Atom Probe Tomography: Analysis at the Atomoc Level. New York: Kluwer Academic/ Plenum.

Murch GE, Bruff CM and Mehrer H. 1990. Chemical diffusion tables. Landolt-Börnstein - Group III Condensed Matter: SpringerMaterials.

Noble B and Bray SE. 1998. On the α(Al)/δ′(Al3Li) metastable solvus in aluminium–lithium alloys. Acta Materialia 46(17): p. 6163-6171.

Nozato R and Nakai G. 1977. Thermal Analysis of Precipitation in Al - Li Alloys, Transactions of the Japan Institute of Metals 18(10): 679-689.

Okuda H, Tanaka M, Osamura K and Anemiya Y. 1993. Synchrotron-radiation small-angle scattering measurements of the reversion process of δ′ precipitates in Al-8.1%Li binary alloy. Acta Metallurgica et Materialia 41(6): 1733-1738.

Sanders TH, Starke Jr. 1984. Aluminum-lithium alloys: in proceding of second imternational aluminum-lithium conference. The metalluurgical socoety of AIME. Warrendale Pennsylvania.

Shariqa A, Mutasa M, Wedderhoffa K, Kleinb C, Hortenbachc H, Teichertc S, Küchera P and Gerstld SSA. 2009. Investigations of field-evaporated end forms in voltage- and laser-pulsed atom probe tomography. Ultramicroscopy 109: 472–479.

Vaumousse D, Cerezo A and Warren PJ. 2003. *A* procedure for quantification of precipitate microstructures from three-dimensional atom probe data. Ultramicroscopy 95(0): 215-221.

Vella A, Vurpillot F, Gault B, Menand A and Deconihout B. 2006. Evidence of field evaporation assisted by nonlinear optical rectification induced by ultrafast laser. Physical Review B 73(16): 165416.

Vurpillot F, Houard J, Vella A and Deconihout B. 2009. Thermal response of a field emitter subjected to ultra-fast laser illumination. Journal of Applied Physics: D 42(125502).

Williams DB and Edington JW. 1975. The Precipitation of δ′(Al3Li) in Dilute Aluminum− Lithium Alloys. Metal Science and Heat Treatment 9(1): 529-532.

.